# New terahertz dielectric spectroscopy for the study aqueous solutions

Deepu K. George, Ali Charkhesht, and N. Q. Vinh[a]

Department of Physics, Virginia Tech, Virginia 24061, USA

ABSTRACT

We present a development of a high precision, tunable far-infrared (terahertz) frequency-domain dielectric spectrometer for studying the dynamics of biomolecules in aqueous solutions in the gigahertz-to-terahertz frequency. As a first application we report on the measurement of the absorption and refractive index for liquid water in the frequency range from 5 GHz to 1.12 THz (0.17 to 37.36 cm$^{-1}$ or 0.268 to 60 mm). The system provides a coherent radiation source with a power up to 20 mW in the gigahertz-to-terahertz region. The power signal-to-noise ratio of our instrument reaches $10^{15}$ and the system achieves a spectral resolution of less than 100 Hz. The temperature of samples can be controlled precisely with an error bars of ±0.02 ºC from above 0 ºC to 90 ºC. Given these attributes, our spectrometer provides unique capabilities for the accurate measurement of even very strongly absorbing materials such as aqueous solutions.

a) Electronic email: Vinh@vt.edu

## I. INTRODUCTION

Terahertz frequency radiation provides unique opportunities to probe the picosecond to nanosecond timescale dynamics properties of biomaterials in liquid water[1-3]. The dissolved biomolecules exhibit low-frequency collective vibrational modes corresponding to conformational changes of biomolecules, such as, for example, the twisting and deformation of the DNA double-helix structure that can be probed directly by the terahertz radiation[5]. The low-frequency vibrational modes in hydrated biomolecules have been proposal as being responsible for efficiently directing biochemical reactions and biological energy transport. Nonetheless, detailed knowledge of the structure and dynamics of aqueous liquid remains to be an outstanding problem in the physical and biological sciences. Our understanding of the translational and rotational diffusion of water molecules and larger-scale rearrangements of its hydrogen-bonding network appears to be incomplete as significant debates regarding the vibrational and relaxational responses of water molecules at the femtosecond to picosecond timescales[1, 3, 7-9]. Unlike infrared and Raman spectroscopies, which are sensitive to femtosecond-scale intramolecular dynamics (i.e., bond vibrations), spectroscopy in the terahertz regime is sensitive to their picosecond intermolecular dynamics (i.e., molecular rotations associated with hydrogen bond breaking) as well as internal motions of solvated biomolecules. Spectroscopy in this regime thus provides a new window with which to study the dynamics of hydrated biomolecules, bulk solvent, and the water in the hydration shells of dissolved biomolecules. Unfortunately, daunting technical limitations associated with this frequency regime, including the extremely strong absorbance of water and other similarly polar materials and often severe interference artifacts, have reduced the precision of prior terahertz spectroscopy studies, limiting our ability to characterize any but the largest-scale, most strongly interacting dynamic modes.

On the optical side of the electrometric spectrum, a number of techniques has been reported for the absorption as well as refractive spectroscopy in the terahertz region. Fourier transform spectrometer (FTS) or Michelson interferometer is a popular technique for broad frequency applications in the infrared to mid-infrared frequencies. This technique obtains information on both the refractive index and the absorption properties of the sample. The technique employs a broadband radiation source which can cover the far-infrared or the terahertz region. However, the power of a typical light source at the terahertz frequencies is very week, limiting the signal-to-noise of the technique in this region. Liquid water is highly absorbing in the terahertz frequencies, thus measurements have been done with a thin layer of water in the transmission[4,6] or in the reflection[10,11] configurations. In order to increase the signal-to-noise of the method at the terahertz region, the measurements have been employed far-infrared gas lasers containing methanol or methyl iodide at low pressures with powers of several mW.[12,13] The method is limited to a number of discrete wavelengths depending on the gas (typically, a few laser wavelengths from 95 μm to 1258.3μm) due to discrete rotational transitions.

Recently, the absorption of liquid water using Free-Electron Lasers,[14] synchrotrons,[15] and germanium laser[18] with high power of radiation in the terahertz frequencies have been reported. However, the lasers only provide a limited tunability over a short range of frequency and only the absorbance (not the refractive index) of the liquid water was extracted from the measurements. In aqueous solution experiments such as protein solutions, the absorbance of solvated proteins was extracted from the measured spectra of protein solutions and the relevant buffer blanks by assuming that the absorbance of the mixture is the weighted average absorbance of its components, an assumption that appears unfounded. Specifically, the true dielectric response (equivalently the absorbance and refractive index) of a mixture is not the weighted sum of the dielectric responses of its components.[9,16,19-21] Moreover, this problem is particularly acute when the refractive index of the solvent changes very rapidly with frequency, as it does for aqueous solutions below 1 THz.

Terahertz time-domain spectroscopy[16, 17, 22-24] pumped by a femtosecond laser pulse generates a fast current pulse (~1 ps) in a dipole antenna on low-temperature grown GaAs. This leads to the emission of electromagnetic pulse. The obtained waveform is then Fourier transformed to the frequency domain power spectrum in the terahertz range from 200 GHz to several THz, depending on the material, the structure of the antenna, and the length of the fs pulse. It is a fast method with good reproducibility and it yields information on the real and imaginary components (or the absorption and refractive index) of materials. The disadvantage is the steep power roll-off leading to low signal-to-noise ratio for higher frequencies in the terahertz region.

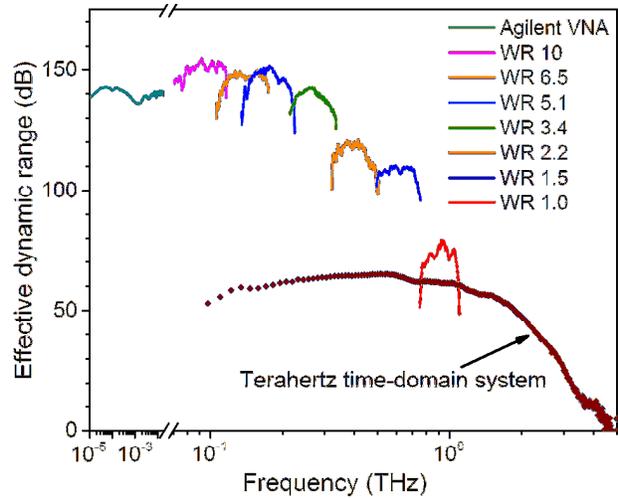

**Figure 1:** Dynamic range of our gigahertz-to-terahertz frequency-domain spectrometer (Agilent Vector Network Analyzer and VDI-frequency extenders from WR-10 to WR-1.0 systems) compared with the dynamic range of a typical terahertz time-domain system.

On the microwave side of the spectrum, the dielectric spectroscopy has been employed to provide information of the microstructure and molecular dynamics of liquid system, especially for aqueous solution. Barthel J. *et al*.[25] and Kaatze *et al*.[26] used the microwave waveguide interferometer in the transmitted configuration and coaxial-line reflection probe, reflectively, to obtain the dielectric relaxation spectra of water up to 89 GHz. The techniques measure simultaneously the absorption and refractive index of solution samples in a broad frequency range but is limited to the GHz frequency. In general the main problem in the terahertz spectroscopy is the lack of high power, high dynamics range, high resolution and a large tunable frequency of radiation sources that limit us the study the conformational dynamics of biomolecules in the living environment.

Here we introduce our terahertz frequency-domain spectrometer, which combines the important elements of high dynamics range with power up to 20 mW, tunable frequency, broadband emission in a table top experiment, demonstrating the possibility to accurately measure absorption as well as refractive index of aqueous solutions. We demonstrate that the terahertz frequency-domain spectrometer is a powerful tool for the dielectric spectroscopy in the gigahertz-to-terahertz frequency. As a first fundamental test sample we have studied pure water. With the high power, we are able to measure thick layers up to 3 mm of liquid water. The large dynamics range of our system minimizes problems associated with multiple reflections of the incident light (standing waves, etalon effect) by using a shortest thickness of the sample. We have measured the absorption and refractive index of water and aqueous solutions over the three order of magnitude range 5.0 GHz – 1.12 THz with high accuracy of temperature at 20.00 (±0.02) °C. The system closes the gap between microwave region and the mid-infrared which is well established for the FTIR measurements.

## II. EXPERIMENT SETUP

In an effort to improve our understanding of the picosecond dynamics of water and solvated molecules, we have built a gigahertz-to-terahertz frequency-domain dielectric spectrometer that supports the simultaneous measurements of absorbance and refractive index of solutions over the spectral range from 5 GHz to 1.12 THz (0.17 to 37.36 cm$^{-1}$ or 0.268 to 60 mm). The signal-to-noise and spectral resolution of this device are significantly improved relative to any previous state-of-the-art instruments. For example, while the power signal-to-noise ratio of a commercial terahertz time-domain spectrometer is just $10^6$ and its spectral resolution is several gigahertz, the power signal-to-noise ratio of our instrument reaches an unprecedented value of $10^{15}$ and the system achieves a spectral resolution of less than 100 Hz (Fig. 1). The system provides a coherent radiation source with a power up to 20 mW in the gigahertz-to-terahertz region. Given these attributes, our spectrometer provides unique capabilities for the accurate measurement of even very strongly absorbing materials such as aqueous solutions[9, 19].

Our spectrometer consists of a commercial Vector Network Analyzer (VNA) from Agilent, the N5225A PNA, which covers the frequency range from 10 MHz to 50 GHz, and frequency multipliers and the matched harmonic detectors for terahertz radiation, which are developed by Virginia Diodes, Inc. (Charlottesville, VA). Detailed information of the vector network analyzer frequency extension modules and the mixer process can be obtained elsewhere.[27, 28] The principle of the frequency extender terahertz modules is shown in Fig. 2. Rather than using optical sources and mixing down the frequency to access the terahertz range, conceptually one can begin with a lower frequency, i.e., microwave sources and multiply the frequency into the terahertz region. One method of fabricating microwave frequency multipliers is to use Schottky diode based components.[28] For example, Virginia Diodes has fabricated autarkic (meaning independent or free from external control or constraint) transmitter and receiver modules. These transmitter and receiver modules (schematic shown) in Fig. 2 were used by Jastrow *et al*.[27] to demonstrate the feasibility of terahertz communications. The transmitter module allows to up-convert an arbitrary signal from a vector network analyzer in the frequency range between 10 MHz and 50 GHz to the terahertz frequency region and transmit it with a rectangular-to-circular horn antenna into free-space. Specifically, in the Fig. 2, the RF (radio frequency) input from a VNA with frequency range from 24.444 to 36.667 GHz enters the WR3.4 frequency extension modules for up conversion frequency by nine times to 220 to 330 GHz. At the receiver module a second horn antenna serves to receive the signal after a sample and feeds it into the mixer for down conversion. The transmitted as well as received signals mix with a Local Oscillator (LO) from the VNA in a subharmonic mixer. The resulting Intermediate Frequency (IF) signal is then detected by the VNA for the reference power and phase shift of the radiation source as well as the measurement signals after the sample, respectively. In this case, the IF signals are the difference between up-converted signals of RF and LO signals at 0.279 GHz. Our terahertz sources from Virginia Diodes for the WR 3.4 frequency extension module produce several milliwatts of power at 300 GHz. Our system composed of the frequency extenders and the matched harmonic receivers was developed by Virginia Diodes, Inc. including WR 10, WR 6.5, WR5.1, WR 3.4, WR 2.2, WR 1.5 and WR 1.0 to cover the frequency range from 60 GHz to 1.12 THz. The power signal-to-noise ratio of the instrument reaches $10^{15}$ with a spectral resolution of less than 100 Hz (Fig. 1). Moreover, the performance and collecting signal of the system are very fast in a real-time of 35 µs per one frequency. It is allows us to measure the dynamics response of a system as a function of time.

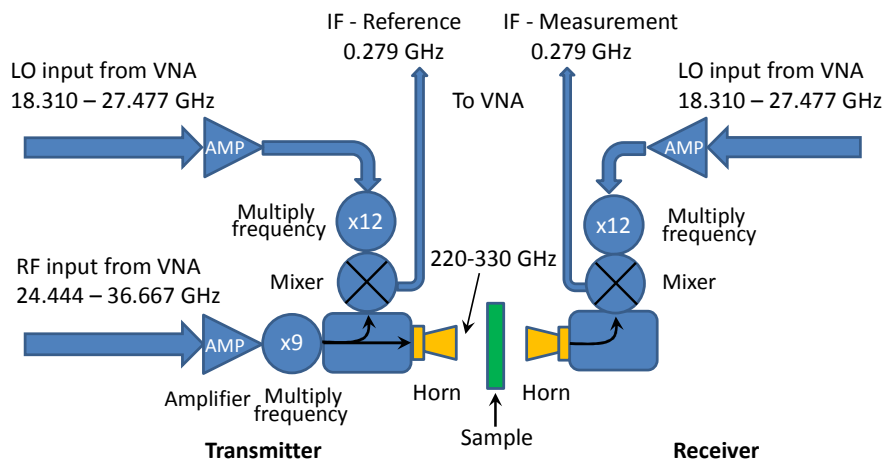

**Figure 2:** Block diagram of the WR3.4 (220 - 300 GHz) transmitter and receiver frequency extender modules. The microwave source from Agilent Vector Network Analyzer is extended via custom Virginia Diode frequency extenders to cover up to 1.12 THz.

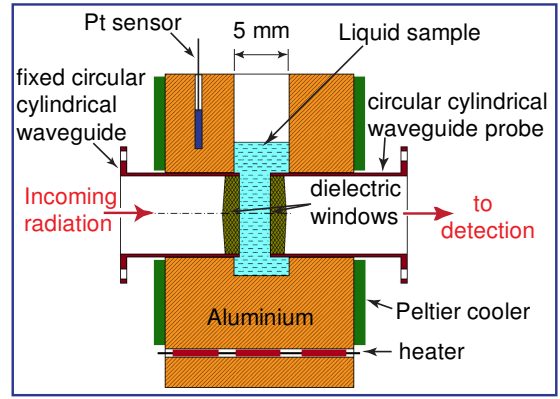

**Figure 3:** A variable path-length sample cell measures how absorbance and refractive index change with changing path-length of sample cell.

We have employed a variable path-length cell setup[16, 29] consisting of two parallel windows inside an aluminum cell, one immobile and the other mounted on an ultra-precise linear translation stage (relative accuracy of 50 nm) (Fig. 3). We use thin polyethylene sheets of 2 µm thickness for the two parallel windows to cover the circular side of the horn antenna with a diameter of 2.85 mm. The thin parallel windows avoid the etalon effect and allow the transmission of terahertz radiation without losses. The metal cell minimizes the leakage of stray radiation. The cell is mounted on Peltier temperature control plates, allowing precise control of the temperature of the sample. The absorbance and refractive index of water are extremely sensitive to temperature, and thus all experiments are carried out with an accuracy of 0.02 °C. The thickness of liquid samples or the distance between the two windows, as the sample path-length, is adjusted using the ultra-precise linear stage. At each frequency, we examine an average of 200 different path-lengths, with increments ranging from 0.1 to 20 µm, depending on the absorption strength of the sample. The temperature of the cell can be controlled precisely from above 0 °C to 90 °C. To mitigate problems associated with multiple reflections of the incident light (standing waves, etalon effect), the thickness of our shortest path-length was selected to be long enough to insure strong attenuation of the incident radiation (transmission <$10^{-2}$).

III. DATA EVALUATION

*Absorption and refractive index measurements.*

Using the above-described spectrometer and sample cell, we have measured the change of intensity and phase in aqueous samples as functions of path-length (Fig. 4). The absorption process of the terahertz radiation passing through a sample is described by Beer's law:

$$I(l, \nu) = I_0(\nu) \cdot e^{-\alpha(\nu) \cdot l} \quad (1)$$

where $\nu$, $I_0$, $I$, $\alpha(\nu)$ and $l$ are the frequency, the incident intensity, the intensity at the detection of the radiation, the absorption coefficient as a function of radiation frequency, and the thickness of the sample, respectively. When the radiation passes through a material, it will always be attenuated. This can be conveniently taken into account by defining a complex refractive index:

$$n^*(\nu) = n(\nu) - i\kappa(\nu) \quad (2)$$

with the real part, $n(\nu)$, is the refractive index and indicates the phase velocity, while the imaginary part, $\kappa(\nu)$, is called the extinction coefficient and indicates the amount of attenuation when the radiation propagates through the material. The extinction coefficient is related to the absorption coefficient, $\alpha(\nu)$, by

$$\alpha(\nu) = \frac{4\pi \cdot \nu \kappa(\nu)}{c} \quad (3)$$

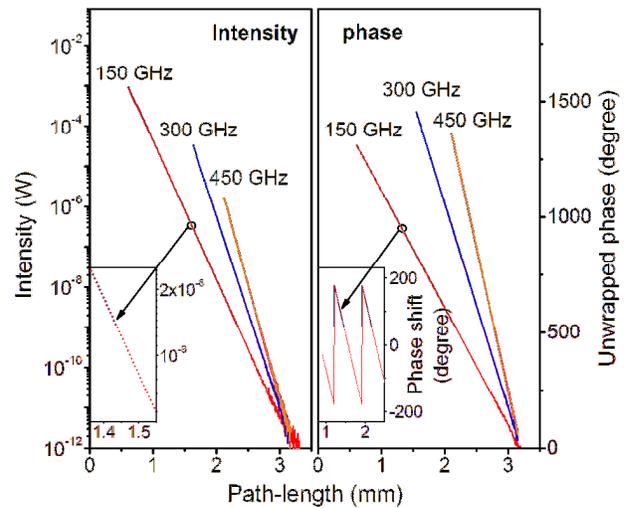

**Figure 4:** The variable path-length sample cell measures the (left) intensity and phase (right) of transmitted terahertz radiation as functions of path-length. The slopes of these lines define the absorbance coefficient and refractive index of water, respectively, without the need for knowledge of the (difficult to obtain) absolute path length or absolute absorbance of our samples. The insets to the figure demonstrate the quality of the measurements. The data in the right inset illustrates the phase shift as a function of sample length.

with $c$ is the speed of light. We have measured the intensity and phase shift of water and aqueous solutions over the three order of magnitude range 5.0 GHz – 1.12 THz as functions of path-length, $l$, at 20.00 (±0.02) °C. The absorption coefficient is determined by the slope of a linear fit of ln $I(l,\nu)$ to the path-length, $l$, without the need for precise knowledge of the sample's absolute absorbance or absolute path-length:

$$\ln I(l, \nu) = \ln I_0(\nu) - \alpha(\nu) \cdot l \quad (4)$$

In parallel, we also fit the observed phase shift $\theta(l,\nu)$ as a linear function of path-length to define the refractive index, $n(\nu)$, of the sample:

$$\theta(l, \nu) = \theta_0(\nu) + \frac{2\pi \cdot \nu n(\nu)}{c} \cdot l \quad (5)$$

where the $\theta_0(\nu)$ is the phase of the reference signal (before the sample). Both properties of liquid water (absorption coefficient and refractive index) are strong functions of frequency, monotonically increasing and decreasing, respectively, with rising frequency over this entire spectral range (Fig. 5).

This method supports the precise determination of absorption coefficients and refractive indexes without the need for precise (and difficult to obtain) measurements of the absolute path-length and the intrinsic optical properties of the sample cell. All experiments were repeated approximately ten times to estimate confidence limits. Using our sensitive setup, we measured precisely the absorption coefficient and refractive index of the strong absorption material, water (Fig. 5), and aqueous solutions at the terahertz frequencies. The red, continuous lines on these two plots in Fig. 5, are water spectra collected with our instrument at 20 °C. Superimposed on these are data collected from the literature,[10, 11, 14, 16, 23, 30, 31] illustrating the vastly improved signal-to-noise and spectral resolution of our instrument.

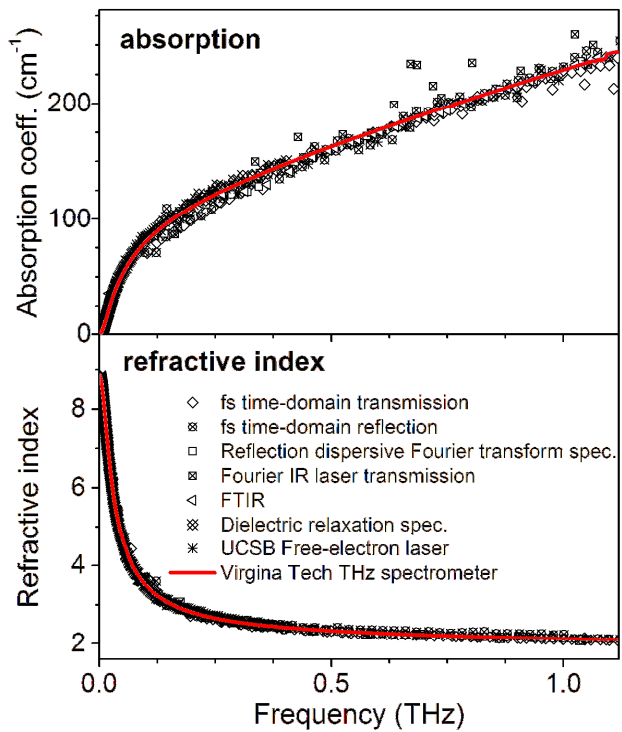

**Figure 5:** The red, continuous lines on these two plots are water spectra collected with our instrument. Superimposed on these are data collected from the literature including measurements using FTIR interferometer ($\nabla$,[4] and ◀[6]), reflection dispersive Fourier transform spectroscopy ($\oplus$[10] and O[11]), far-infrared lasers (▲,[12] ▶,[13]), free-electron laser (●)[14], terahertz time-domain transmission (◊[16, 17], □[22]) and reflection (⊗,[23] Δ[24]) spectroscopies, dielectric relaxation spectroscopy (♦,[25] ■[26]),

*Complex dielectric response of solutions.*

The spectroscopies cover a broadband spectral range from gigahertz to terahertz frequencies that allow us to observe both the relaxational (rotational) and translational processes of waters and biomolecules.

Thus, the dielectric response will provide an entire picture of the dynamics of biomolecules in the living environment. The frequency-dependent complex dielectric response, $\varepsilon^*(\nu) = \varepsilon'(\nu) - i\varepsilon''(\nu)$, is related to the complex refractive index, $n^*(\nu) = n(\nu) - i\kappa(\nu)$, through the relations:[32]

$$\begin{aligned}\varepsilon'(\nu) &= n^2(\nu) - \kappa^2(\nu) = n^2(\nu) - (c\alpha(\nu)/(4\pi\nu))^2 \\ \varepsilon''(\nu) &= 2n(\nu)\cdot\kappa(\nu) = 2n(\nu)c\alpha(\nu)/(4\pi\nu)\end{aligned} \quad (6)$$

From this we can determine the complex dielectric function, $\varepsilon^*(\nu)$, of the water and aqueous solutions, which in turn provides a complete description of the interaction of the solution with the incoming electromagnetic wave. Figure 6 shows the dielectric response from water at 20 °C. Conversely, given the complex dielectric response, $\varepsilon'(\nu)$ and $\varepsilon''(\nu)$, we can determine the absorption, $\alpha(\nu)$, and the refractive index $n(\nu)$:

$$\begin{aligned}n(\nu) &= \left(\frac{\sqrt{\varepsilon'(\nu)^2+\varepsilon''(\nu)^2}+\varepsilon'(\nu)}{2}\right)^{1/2} \\ \alpha(\nu) &= \frac{4\pi\nu}{c}\left(\frac{\sqrt{\varepsilon'(\nu)^2+\varepsilon''(\nu)^2}-\varepsilon'(\nu)}{2}\right)^{1/2}\end{aligned} \quad (7)$$

## IV. DISCUSSION

We have demonstrated that we are now able to qualify the absorption as well as the refractive index of aqueous solutions with the dielectric terahertz spectroscopy. The system reduces the influence of interference and other systematic effects to minimum and provides reliable absolute experimental data for water and aqueous solutions. The absorption coefficient as well as the refractive index of water strongly depend on temperature. We have employed the Peltier system to control the temperature with a high accuracy of 0.02 °C. It is a fast technique that allows us to observe the conformation changes of biomolecules in solutions as a function of time.

Figure 5 shows the result of our measurement of liquid water measured at 20 °C. The absorption coefficient, and refractive index are strong functions of frequency, monotonically increasing and decreasing, respectively, with rising frequency over this entire spectral range. The red, continuous lines collected with our instrument for water spectra are in good agreement with previously reported spectra (Fig. 5). The Zelsmann[4] and Zoidis et al.[6] recorded the water spectra with a FTIR interferometer in the range from 8 to 450 cm$^{-1}$ (239 GHz to 13.5 THz) using the transmission configuration. Afsar et al.[10] and Hasted et al.[11] employed a reflection dispersive Fourier transform spectroscopy to measure to absorption and refractive index of liquid water in the spectral range between 4 and 450 cm$^{-1}$ (120 GHz to 13.5 THz). Using far-infrared gas lasers with powers of several mW, Vij et al.[12] and Simpson et al.[13] reported the absorption coefficients of liquid water in a few frequencies in the terahertz region. Xu et al.[14] measured the absorption coefficient of liquid water between 0.3 to 3.75 THz with free-electron lasers. Schmuttenmaer et al.[16, 17] and Yada et al.[22] reported the absorption coefficient and refractive index with a terahertz time-domain system in the transmission configuration in the range from 2.0 to 60 cm$^{-1}$ (60 GHz to 1.8 THz) using a variable path-length sample cell. The absorption coefficient was determined by the slope of a linear regression fit of the detected intensity versus the path-length at room temperature. Thrane, Ronne et al.[23, 24] collected the terahertz spectrum of liquid water at 292 K using a terahertz time-domain system in the reflection configuration in the range from 3.0 to 33 cm$^{-1}$ (90 GHz to 1.0 THz). Barthel J. et al.[25] and Kaatze et al.[26] used the microwave waveguide interferometer in the transmitted configuration and coaxial-line reflection probe, reflectively, to obtain the dielectric relaxation spectra of water up to 89 GHz. Superimposed on these are data collected from the literature,[10, 11, 13, 14, 16, 23, 30, 31] illustrating the significantly improved signal-to-noise and spectral resolution of our terahertz spectrometer.

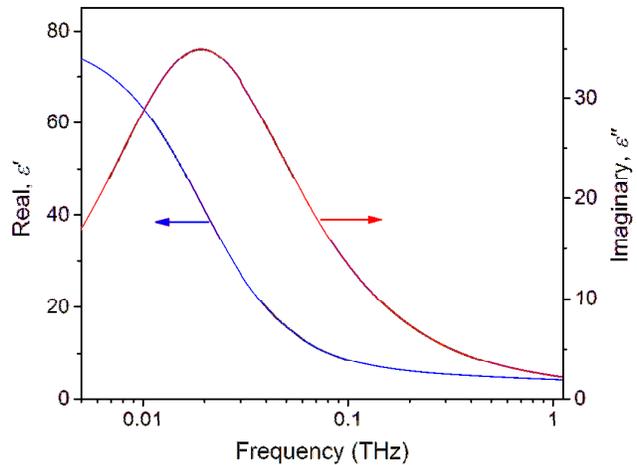

**Figure 6:** The dielectric response from water at 20 °C is converted from the absorption coefficient and refractive index measurements.

In summary, we have demonstrated a new simple method to obtain terahertz spectra of highly absorption polar liquid with a high precision, high dynamics range, high resolution and a large frequency range from gigahertz-to-terahertz region. The terahertz frequency-domain dielectric spectroscopy applied to liquid water obtained a good agreement with previous measurements. Using this setup we have been able to determine the absorption coefficient and the refractive index of water as well as the aqueous biological solutions in the range between 5 GHz and 1.12 THz with high precision.

V. ACKNOWLEGMENTS

This work was supported by the Institute of Critical Technology and Applied Sciences (ICTAS) at Virginia Tech.